\documentstyle[amssymb,aps,epsf,twocolumn]{revtex}

\def\be{\begin{equation}}
\def\ee{\end{equation}}
\def\bea{\begin{eqnarray}}
\def\eea{\end{eqnarray}}
\def\nn{\nonumber\\}

\begin{document} 
\twocolumn[\hsize\textwidth\columnwidth\hsize\csname
@twocolumnfalse\endcsname

\title{Metamagnetism in one dimensional systems with edge sharing CuO polihedra}
\author{A. A. Aligia}
\address{Comisi\'{o}n Nacional de Energ{\'{\i }}a At\'{o}mica,\\
Centro At\'{o}mico Bariloche and Instituto Balseiro, 8400 S.C. de Bariloche,%
\\
Argentina}
\maketitle

\begin{abstract}
We study a Heisenberg chain with nearest-neighbor (NN) $J_{1}$ and next-NN $%
J_{2}$ exchange interactions with anisotropies $\Delta _{1}$ and $\Delta
_{2} $ respectively. We investigate by analytical and numerical methods the
region of parameters for which there is a jump in the magnetization $M$ as a
function of magnetic field $B$. Some materials with edge sharing CuO
polihedra are candidates to show an abrupt change in $M(B)$.
\end{abstract}

\vskip2pc]

\narrowtext

The magnetization as a function of applied magnetic field in several
materials \cite{ito,lel,eck} shows a discontinuity or very rapid increase at
a certain field $B_{c}$. Gerhardt {\it et al. }have shown that for certain
parameters, a magnetization jump is also present in the spin-1/2 XXZ chain
with NN and next-NN exchange coupling (keeping $\Delta _{1}=\Delta
_{2}=\Delta $) \cite{ger}:

\begin{eqnarray}
H &=&\sum_{i}[J_{1}(S_{i}^{x}S_{i+1}^{x}+S_{i}^{y}S_{i+1}^{y}+\Delta
_{1}S_{i}^{z}S_{i+1}^{z})  \nonumber \\
&&+\sum_{i}J_{2}(S_{i}^{x}S_{i+2}^{x}+S_{i}^{y}S_{i+2}^{y}+\Delta
_{2}S_{i}^{z}S_{i+2}^{z})]   \nn 
-B\sum_{i}S_{i}^{z},  \label{ham}
\end{eqnarray}

For a metamagnetic transition to occur at very low temperatures, the
zero-field ground-state energy per site $E$ as a function of the
magnetization $M=\sum_{i}S_{i}^{z}/L$ ($L$ is the number of sites), should
satisfy two conditions: I) $\partial ^{2}E/\partial M^{2}<0$ in a finite
interval of values of $M$. Then one can draw a straight line $E^{\prime }(M)$
which is tangent to $E(M)$ in at least two points (the Maxwell construction,
see Fig. 1 (a)) in such a way that $E(M)\geq E^{\prime }(M)$ for $M_{1}\leq
M\leq M_{2}$. II) $E(M_{2})>E(M_{1})$. If these two conditions are
satisfied, $M$ jumps from $M_{1}$ to $M_{2}$ at the critical field $%
B_{c}=[E(M_{2})-E(M_{1})]/(M_{2}-M_{1})$.

From the general behavior of $E(M)$, Gerhardt {\it et al. }have found that
when metamagnetism exists, $M_{2}=1/2$ and the condition II ceases to be
satisfied when $M_{1}=0$. More precisely, from their finite-size results for 
$E(M,\alpha ,\Delta )$, with $\alpha =J_{2}/J_{1}$, they obtained a critical
value of $\Delta $ ($\Delta _{f}(\alpha )$) from the equation $E(0,\alpha
,\Delta _{f})=E(1/2,\alpha ,\Delta _{f})$. For $\Delta <\Delta _{f}$ the
system is ferromagnetic at $B=0$. Another critical value $\Delta _{a}(\alpha
)$ was obtained from the condition $\partial ^{2}E/\partial M^{2}|_{M=1/2}=0$%
. For $\Delta >\Delta _{a}$ the curvature $\partial ^{2}E/\partial M^{2}$ is
positive for all $M$. The discretized $\partial ^{2}E/\partial
M^{2}|_{M=1/2}=0$ has some finite-size effects \cite{ger}. From the
numerical solution of the problem of two spin excitations on the
ferromagnetic state for $L\rightarrow \infty$, more accurate values of $%
\Delta _{a}(\alpha )$ were obtained recently for $\alpha \leq 1/2$ \cite{hir}%
. In the region of the $(\alpha ,\Delta )$ plane where $\Delta _{f}(\alpha
)<\Delta <\Delta _{a}(\alpha )$ a metamagnetic transition occurs in the
model \cite{ger,hir}.

We have studied the two-magnon problem for generic values of $\Delta _{1}$
and $\Delta _{2}$, and found analytical results for the condition $\partial
^{2}E/\partial M^{2}|_{M=1/2}=0$ if $\alpha \leq 0.75$. When $\Delta
_{1}=\Delta _{2}=\Delta $, in the region $\alpha \leq 1/4$, the function $%
\Delta _{f}(\alpha )$ can be accurately approximated by:

\begin{eqnarray*}
\Delta _{f} &=&-1+2\sum_{i=1}^{4}\alpha ^{i}+6\alpha ^{5}+O(\alpha ^{6})%
\text{, if }\alpha \leq 0.2 \\
\Delta _{f} &=&\frac{1}{4}(-5+\sqrt{17})-0.462\sqrt{1-4\alpha }\text{, }%
0.2\leq \alpha \leq \frac{1}{4}.
\end{eqnarray*}
For $1/4\leq \alpha \leq 1/2$, although the algebra is more involved, the
exact result is simpler:

\[
\Delta _{f}=-b+\sqrt{b^{2}-2\alpha }\text{, with }b=\alpha +\frac{1}{2}+%
\frac{1}{8\alpha }. 
\]
Finally in the region $1/2\leq \alpha \leq 0.75$, $\Delta _{f}(\alpha )$ is
very flat. Near $\alpha =1/2$ it can be approximated as $\Delta _{f}=-\frac{1%
}{2}+0.309(x-\frac{1}{2})^{2}$. These results show that metamagnetism is not
possible if $\Delta >(-5+\sqrt{17})/4=-0.219$. Unfortunately, such a large
anisotropy of $J_{2}$ seems unrealistic. Instead, $\Delta _{1}=-1$
corresponds to isotropic ferromagnetic $J_{1}$, since a rotation of every
second spin in $\pi $ around the $z$ axis changes the sign of the $x$ and $y$
components of $J_{1}$.

The main purpose of this work is to extend the previous results to negative $%
\Delta _{1}$ and positive $\Delta _{2}$. Since it is expected that the
parameters for several copper oxides containing edge sharing Cu-O chains lie
near the isotropic limit $\Delta _{1}=-1$, $\Delta _{2}=1$,\cite{miz} we
consider this limit in what follows. From numerical diagonalization of 20
sites, we obtain that spontaneous ferromagnetism does not take place for $%
\alpha >1/4$. If in addition $\alpha \leq 0.7$, there is a bound state in
the two-magnon problem at wave vector $q_{2}=2q_{1}$, where $q_{1}=\pm
\arccos [-1/(4\alpha )]$ are the wave vectors of the one-magnon states of
lowest energy. For $\alpha >0.7$, there might be a two-magnon bound state
with $q_{2}\neq 2q_{1}$, but we have not studied this alternative because it
seems not possible to solve the problem analytically for large $\alpha $.
Thus, we expect a jump in $M(B)$ for $1/4<\alpha \leq \alpha _{c}$ with $%
\alpha _{c}\geq 0.7$.

In Fig. 1(a) we show $E(M)$ for a chain of $L=20$ sites with periodic
boundary conditions for $\alpha =0.425$, chosen in such a way that $%
q_{1}=\pm 7\pi /10$ are allowed wave vectors of the finite chain. For other
values of $\alpha $, one might obtain a numerical negative $\partial
^{2}E/\partial M^{2}|_{M=1/2}$ because of frustration effects which increase 
$E(M-1/L)$. In spite of this precaution, the results show a significant
even-odd effect: the energies for odd (even) total spin $S=|M|L$ seem to be
shifted to higher (lower) energies. If this effect persists in the
thermodynamic limit (keeping $L$ even) states with odd $S$ become irrelevant
(because they do not minimize $E-MB$ for any $B$) and a bound state in the
two-magnon problem does not necessarily imply $\partial ^{2}E/\partial
M^{2}|_{M=1/2}<0$. From $E(S/L)$ for the three highest even $S$ with $L=28$,
minimized with respect to the optimum twisted boundary conditions to allow
for incommensurate wave vectors \cite{ali}, we obtain a very small curvature
which is negative for $\alpha <\alpha _{c}=0.359$ but positive for $\alpha
>\alpha _{c}$. If $\alpha _{c}$ remains finite in the thermodynamic limit, $%
M(B)$ would increase abruptly for $\alpha >\alpha _{c}$, but without showing
a true jump. While this difference is hard to distinguish experimentally, it
would be of interest to calculate $\partial ^{2}E/\partial M^{2}|_{M=1/2}$
using larger clusters.

To obtain a continuous curve $E(M)$ from which $M(B)$ can be derived, we
have fitted the eleven numerical values represented in Fig. 1(a) by a
polynomial of even powers of $M$ up to $M^{10}$. This function satisfies the
physical condition $E(M)=E(-M)$ and has six fitting parameters (nearly half
of the number of points to be fitted, to average the even-odd effect). The
resulting $B=\partial E/\partial M$ is represented in Fig. 1(b). At a
critical field $B_{c}=0.192J_{1}$, the magnetization jumps from $M_{1}=0.347$
to $M_{2}=1/2$. While the numerical values of $B_{c}$ and particularly $%
M_{1} $ depend on the particular fitting procedure used and the size of the
system, the general shape of $M(B)$ is robust: around $B/J_{1}=0.19\pm 0.01$
there is a sudden increase of $M$ from $\sim 0.25$ to 1/2. While the shape
of $M(B)$ does not depend very much on $\alpha >1/2$, $B_c$ increases
strongly with $\alpha$.

To conclude, while the existence of a real jump in $M(B)$ requires a study
of larger clusters, we have shown that the magnetization of the model for
parameters appropriate to edge-sharing Cu-O chains has a sudden increase at
a magnetic field $B_{c}$. Using parameters calculated for La$_{6}$Ca$_{8}$Cu$%
_{24}$O$_{41}$, \cite{miz} and assuming a gyromagnetic factor $g=2$ we
obtain $B_{c}\simeq 14$ Tesla.

I am grateful to F.H.L. E\ss ler, Ana L\'{o}pez and C.D. Batista for
important discussions. I am partially supported by CONICET. This work was
sponsored by PICT 03-00121-02153 of ANPCyT and PIP 4952/96 of CONICET.

\begin{figure}
\narrowtext
\epsfxsize=3.5truein
\vbox{\hskip 0.05truein \epsffile{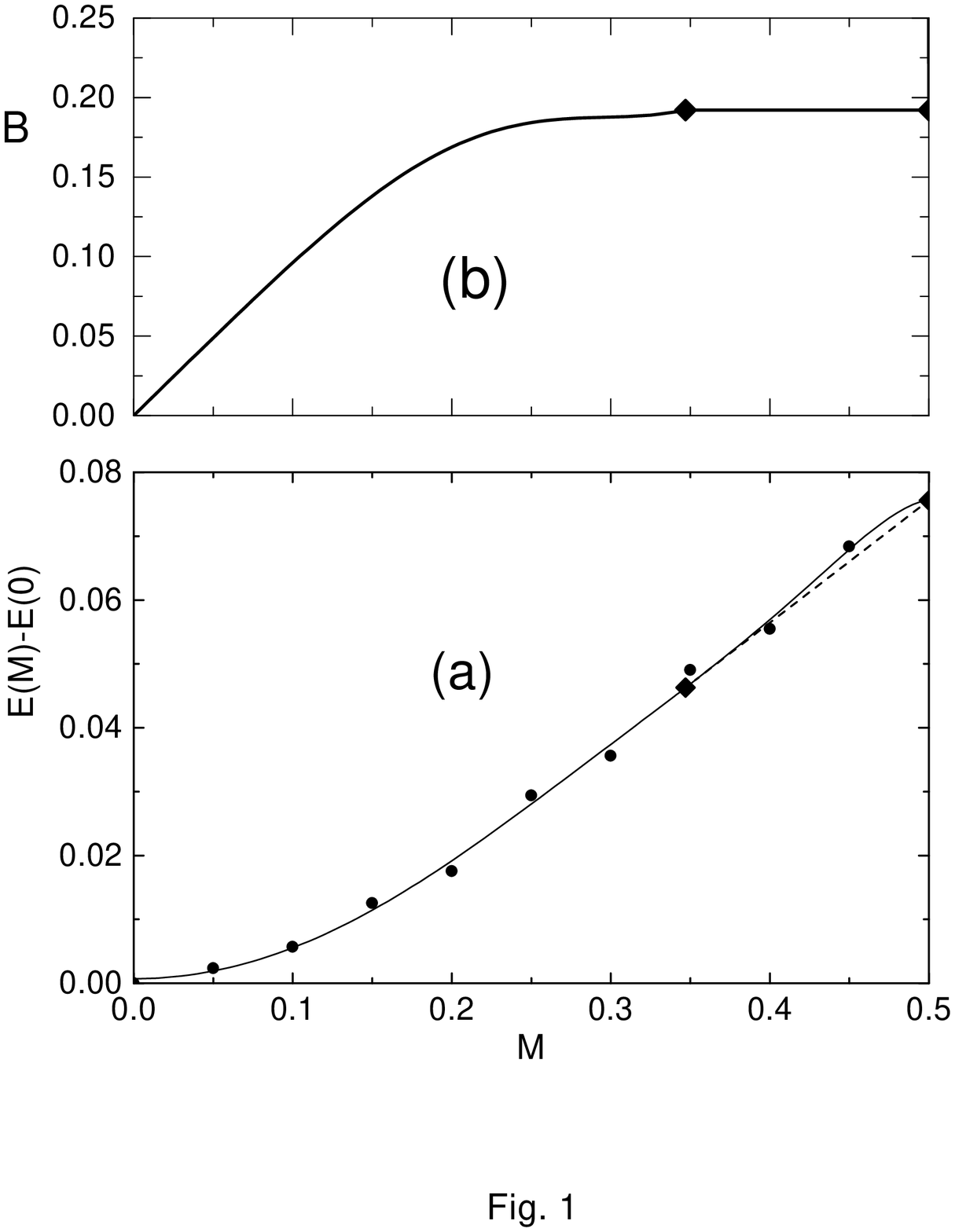}}
\medskip
\caption{Energy per site as a function of total spin per site for a
chain of 20 sites (solid circles). The full line is a fit (see text). Dashed
line and diamonds correspond to the Maxwell construction. (b) Magnetic field
as a function of the magnetization. Parameters are $J_{1}=1$, $J_{2}=0.425$, 
$\Delta _{1}=-1$ and $\Delta _{2}=1$.}
\end{figure}

\end{document}